\documentclass[preprint2]{aastex}

\def\be{\begin{equation}}
\def\ee{\end{equation}}

\begin{document}

\title{Pulsar Magnetosphere} 

\author{Andrei Gruzinov}

\affil{CCPP, Physics Department, New York University, 4 Washington Place, New York, NY 10003}

\begin{abstract}

We argue that existing models of the ideal pulsar magnetosphere are incorrect because of the improper treatment of the singular current layer outside the light cylinder. We simulated the axisymmetric pulsar magnetosphere in the Force-Free limit of Strong-Field Electrodynamics. It turns out that even in the Force-Free limit, some field lines enter the singular current layer which lies  beyond the light cylinder in the equatorial plane. As a result: (i) about 10\% of the Poynting flux is dissipated between 1 and 1.5 light cylinder radii, (ii) there is no singular current layer within the light cylinder.

~~

\end{abstract}

\section{The pulsar magnetosphere problem} 

Consider a magnetized ideally conducting ball rotating in vacuum. If the ball rotates fast enough, is large enough, and magnetized enough, it will create charges around itself. This can happen even before the Schwinger field is reached, by the tree-level QED processes, and also by pulling charged particles from the surface. Thus, the spinning magnetized conducting ball creates a magnetosphere around itself. It is thought that neutron star magnetospheres are described by this model, and the infinitely rich pulsar emission phenomenology somehow follows. Then, to understand pulsar emission, one should first understand the large-scale structure of the pulsar magnetosphere. 

\begin{figure}
\plotone{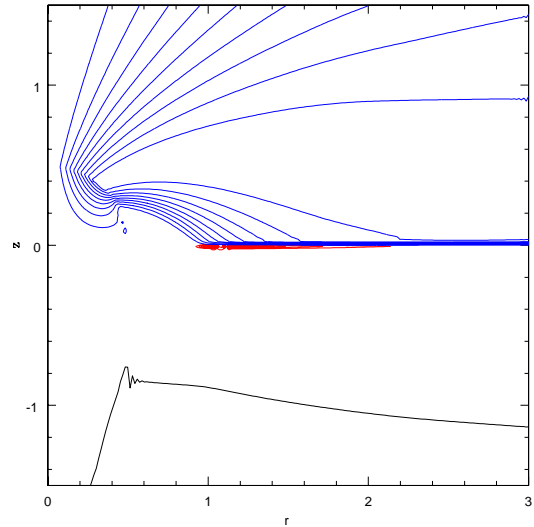}
\caption{Blue: poloidal current, $rB_\theta $, above the equator. Red: dissipation rate, ${\bf E}\cdot {\bf j}$, below the equator, outside the star. Black curve: Poynting flux through the cylindrical surface, $\propto \int rdz(E_\theta B_z-E_zB_\theta )$, as a function of radius. Simulation parameters (in pulsar units, $\Omega =c=1$): $r_s=0.5$, $r_m=3$, $z_m=4.5$, $\sigma = \sigma _s=333$, $\eta =1.5\times 10^{-3}$.}
\end{figure}

The ideal pulsar magnetosphere was ``drawn'' by Goldreich and Julian (1969) and calculated by Contopoulos, Kazanas and Fendt (1999). These calculations were improved by Gruzinov (2005) who also gave the pulsar power formula and some exact results for the axisymmetric pulsar. Spitkovsky (2006) did the non-axisymmetric case. 

All these results, we think, are incorrect. In reality, a fair fraction of the magnetic field lines enters the singular current layer outside the light cylinder. As a result (for axisymmetric pulsar):

1. Some 10\% of the Poynting flux is damped between 1 and 1.5 light cylinder radii.

2. There is no singular current layer within the light cylinder.

The error of all existing calculations comes from the improper treatment of the singular current layer outside the light cylinder. Here we describe simulations of the pulsar magnetosphere which resolve all singularities -- we don't use any boundary conditions, not even at the surface of the star.

Our numerical simulations are done in the FFE (Force-Free Electrodynamics) limit of the SFE (Strong-Field Electrodynamics). We describe FFE and SFE in the next section. In \S3, we describe the simulations.

\section{FFE and SFE} 

Both FFE and SFE are plasma physics models which describe the plasma implicitly (Gruzinov 2008). Namely, one solves the Maxwell equations
\be
\partial _\nu F^{\mu \nu}=-j^\mu,
\ee
or in the 3+1 split,
\be
\partial _t{\bf B} =-\nabla \times {\bf E},~~~ \partial _t{\bf E}=\nabla \times {\bf B}-{\bf j},
\ee
supplemented by some Ohm's law, which gives ${\bf j}$ {\it in terms of the electromagnetic field only}. 

In FFE, the Ohm's law is
\be
F^{\mu \nu}j_\nu =0,~~ E_0=0.
\ee
Here the scalar $E_0$ is the proper electric field, defined by
\be
B_0^2-E_0^2\equiv {\bf B}^2-{\bf E}^2, ~~~ B_0E_0\equiv {\bf E}\cdot {\bf B}, ~~~E_0\geq 0.
\ee

The physical meaning of the FFE Ohm's law is as follows. For any electromagnetic field, at any event, there is a one-parameter family of {\it good} frames, where ${\bf E}$ is parallel to ${\bf B}$. FFE postulates, that in any {\it good} frame, the electric field vanishes and the current flows along the magnetic field. 

In SFE, the Ohm's law is 
\be
B_0F^{\mu \nu}j_\nu =E_0\tilde{F}^{\mu \nu}j_\nu , ~~~ j_\mu j^\mu =-\sigma^2E_0^2.\ee
Here $\tilde{F}$ is the dual tensor, and $\sigma$ is the conductivity scalar. 

The physical meaning of the SFE Ohm's law is as follows. At each event, the family of {\it good} frames contains the {\it best} frame, where the charge density vanishes, and the current $\sigma E_0$ flows along the common direction of the electric and magnetic fields. If $E_0=0$, the charge density $\rho$ has to move at the speed of light, so that $j^\mu j_\mu=0$. 

In numerical simulations, one uses the 3+1 split, and the FFE Ohm's law becomes
\be
{\bf j}={({\bf B}\cdot \nabla \times {\bf B}-{\bf E}\cdot \nabla \times {\bf E}){\bf B}+(\nabla \cdot {\bf E}){\bf E}\times {\bf B} \over B^2}.
\ee
The SFE Ohm's law is
\be \label{ohm}
{\bf j}={\rho {\bf E}\times {\bf B} +(\rho ^2+\gamma ^2\sigma ^2E_0^2)^{1/2}(B_0{\bf B}+E_0{\bf E})\over B^2+E_0^2},
\ee
where 
\be
\gamma ^2 \equiv {B^2+E_0^2\over B_0^2+E_0^2},~~ \rho \equiv \nabla \cdot {\bf E}
\ee

In the limit of high conductivity, SFE reduces to FFE (Gruzinov 2008). This might seem strange, because SFE postulates that the 4-current is always space-like or null-like, $j^\mu j_\mu \leq 0$, while FFE admits time-like currents, $j^\mu j_\mu \equiv \rho ^2-{\bf j}^2>0.$ But it turns out, that SFE handles the time-like currents by constantly switching the direction of the null-like current, such that the time-averaged current becomes time-like. 

FFE can be applied only to initial electromagnetic fields of special geometry -- with the electric field everywhere smaller than and perpendicular to the magnetic field. SFE applies to arbitrary initial field.

FFE is ideal, the electromagnetic energy is conserved. SFE is {\it semi-ideal}, the electromagnetic energy is non-increasing, but it remains exactly constant for all fields with $E_0=0$.

\section{Numerical simulations} 

In cylindrical coordinates $(r,\theta ,z)$, assuming axisymmetric field, we numerically integrate the following equations
\be
\partial _t{\bf B} =-\nabla \times {\bf E}+\eta \Delta {\bf B},
\ee
\be
\partial _t{\bf E} =\nabla \times {\bf B}-{\bf j}+\eta \Delta {\bf E}.
\ee
The small diffusivity $\eta$ is added for regularization. The equations are solved in a volume
\be
r<r_m,~~ |z|<z_m.
\ee
The boundary conditions at $r_m$ are 
\be
\partial _r{\bf B}=\partial _r{\bf E}=0.
\ee
The boundary conditions at $z_m$ are
\be 
{\bf B}={\bf E}=0.
\ee
The initial conditions are
\be 
{\bf B}={\bf E}=0.
\ee
The Ohm's law outside the star
\be 
r^2+z^2>r_s^2
\ee
is given by eq.(\ref{ohm}), with the field invariants regularization of (Gruzinov 2008). The Ohm's law inside the star is the standard relativistic Ohm's law in a moving medium
\be 
{\bf j}=\sigma _s\gamma _s({\bf E}+{\bf v}\times {\bf B})+{\bf v}(\rho -\sigma _s\gamma _s{\bf E}\cdot {\bf v})+{\bf j}_e.
\ee
Here $\sigma _s$ is the conductivity of the star; ${\bf v}$ is the uniform rotation, and ${\bf j}_e$ is the external current -- both purely toroidal, inside the star:
\be 
v_\theta =\Omega r,~~ j_{e\theta } \propto r;
\ee
$\gamma _s$ is the Lorentz factor of ${\bf v}$. 

The numerical scheme was just the direct primitive discretization of the PDEs. The only subtlety is the time step. Since SFE handles the time-like current regions by permanently switching the sign of $B_0$, one needs to reduce the time step (after saturation of the fields) to get accurate final results. 

We increased the spatial resolution and conductivities $\sigma$ and $\sigma _s$, and decreased the diffusivity $\eta$, until the final magnetosphere outside the star showed clear signs of saturation. The final state is shown in Fig.1. As one can see, the magnetosphere is different from what it was thought to be.\footnote{In fact, the dissipative pulsar magnetosphere simulations of Gruzinov (2008b) agree with the results of Fig.1. However, the numerical scheme of Gruzinov (2008b) was not good enough, and at that time we thought that as the conductivity increases, the dissipative pulsar magnetosphere would approach the standard FFE magnetosphere. We now believe that this assumption is wrong.}

\acknowledgements

I thank Peter Goldreich and Anatoly Spitkovsky for many useful discussions.

\end{document}